\documentclass[aps,twocolumn,amsmath,amssymb,superscriptaddress,notitlepage,longbibliography]{revtex4-1}

\usepackage[colorlinks=true,citecolor=blue,linkcolor=blue,hypertexnames=false]{hyperref}
\usepackage{amsmath}
\usepackage{dsfont}
\usepackage{dcolumn}
\usepackage{hyperref}
\usepackage{tikz}
\usepackage{textcomp}
\usepackage{float}
\usepackage{mwe}
\usepackage{graphicx}
\usepackage{ulem}

\def \k{{\mathbf k}}
\def \r{{\mathbf r}}
\def \A{{\mathbf A}}
\newcommand*{\ket}[1]{|#1\rangle}

\begin{document}
\title{Flat band topology of magic angle graphene on a transition metal dichalcogenide}

\author{Tianle Wang}
\affiliation{Department of Physics, University of California, Berkeley, CA 94720, USA}
\affiliation{Materials Science Division, Lawrence Berkeley National Laboratory, Berkeley, California 94720, USA}
\author{Nick Bultinck}
\affiliation{Department of Physics, University of California, Berkeley, CA 94720, USA}
\affiliation{Department of Physics, Ghent university, 9000 Gent, Belgium}
\author{Michael P. Zaletel}
\affiliation{Department of Physics, University of California, Berkeley, CA 94720, USA}
\affiliation{Materials Science Division, Lawrence Berkeley National Laboratory, Berkeley, California 94720, USA}

\date{\today}
\begin{abstract}
    We consider twisted bilayer graphene on a transition metal dichalcogenide substrate, where proximity-induced spin-orbit coupling significantly alters the eight flat bands which occur near the magic angle. The resulting band structure features a pair of extremely flat bands across most of the mini-Brillouin zone. Further details depend sensitively on the symmetries of the heterostructure; we find semiconducting band structures when all two-fold rotations around in-plane axis are broken, and semi-metallic band structures otherwise.  We calculate the Chern numbers of the different isolated bands, and identify the parameter regimes and filling factors where valley Chern insulators and topological insulators are realized. Interestingly, we find that for realistic values of the proximity-induced terms, it is possible to realize a topological insulator protected by time-reversal symmetry by doping two holes or two electrons per superlattice unit cell into the system. 
\end{abstract}

\maketitle

\section{Introduction}

Recent experimental progress \cite{Cao,Cao2,Yankowitz,Kerelsky,RutgersSTM,efetov,YazdaniSpectroscopic,Efetovscreening,YoungScreening,Choi,Sharpe,YoungAH,Tomarken,CascadeYazdani,CascadeShahal,Harpreet,YazdaniChern,AndreiChern,Andreananosquid,EfetovFragile,VafekLi,PabloNematicity} has maintained continued interest in the study of twisted bilayer graphene (TBG) in the  magic angle regime, where the twist angle between the two graphene layers is approximately one degree.
Near this magic angle, the TBG band spectrum structure contains eight bands near the charge neutrality point with a very small bandwidth. When the Fermi level lies within these flat bands, interactions play an important role and lead to the appearance of correlated insulating states at certain integer fillings. Interestingly, the flat bands were also found to have a subtle but non-trivial form of band topology \cite{Po,Po2,ZouPo,Bernevig,Hejazi,LiuDai,SWAhn}. 

In this work, we consider the band spectrum of magic angle graphene in the presence of a transition metal dichalcogenide (TMD) substrate. For some first experimental results on such TBG-TMD devices, at angles $\theta = 0.79$\textdegree$-0.97$\textdegree\ somewhat below the first magic angle, see Ref. \cite{Harpreet}. The heavy atoms in a TMD substrate are known to introduce significant spin-orbit coupling (SOC) in graphene via the proximity effect. Because of their small bandwidth, the flat bands are expected to be significantly reconstructed by the SOC terms, and the band topology of the SOC bands can potentially be very different from the band topology of the original BM bands. Studying how the flat band topology changes in the presence of SOC is not purely a theoretical exercise, but is also an important step towards understanding the interacting phase diagram. One of the main reasons is that magic angle graphene has a large approximate U$(4)\times$U$(4)$ symmetry \cite{KIVC,KangVafekDMRG}, as a result of which there are many different candidate symmetry-breaking states which are very close in energy, as is seen for example in numerical Hartree-Fock \cite{XieMacDonald,KIVC,Choi,Cea}, density matrix renormalization group \cite{KangVafekDMRG} and quantum Monte Carlo \cite{Fernandes} studies. Because of this close intrinsic competition, small extrinsic effects coming from the substrate can tip the balance between different symmetry-breaking states. For example, the two-fold in-plane rotation symmetry breaking staggered sublattice potential which is induced by an aligned hexagonal Boron-Nitride substrate has already been observed to drastically change the interacting phase diagram of magic angle graphene \cite{Sharpe,YoungAH,Andreananosquid}. This is in agreement with the numerical Hartree-Fock study of Ref. \cite{KIVC}, where it was found that a staggered sublattice potential as small as $10$ meV can change the nature of the ground state at charge neutrality. 

\begin{figure}[h]
    \centering
    \includegraphics[width=8cm]{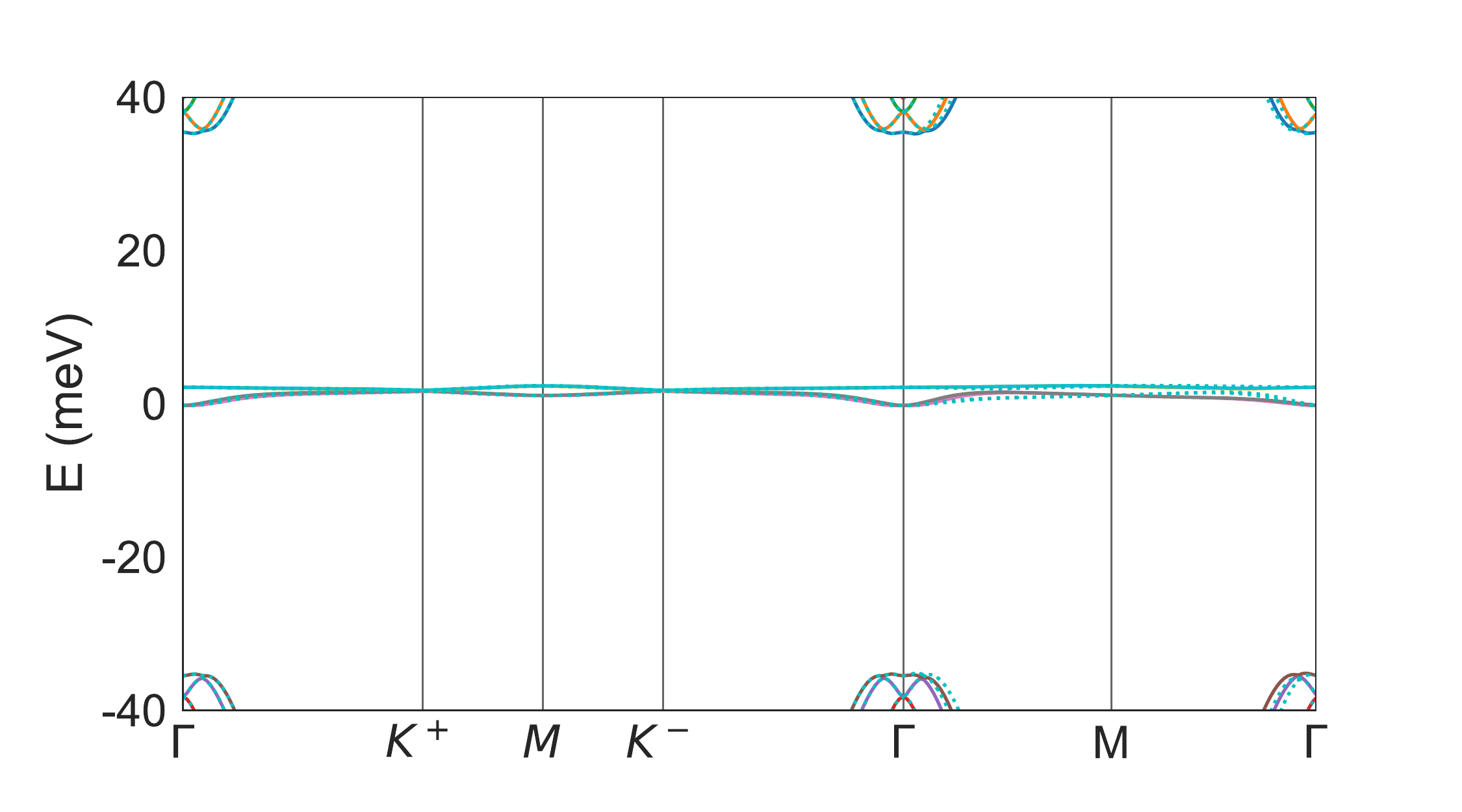}
    \\ (a) \\
    \includegraphics[width=8cm]{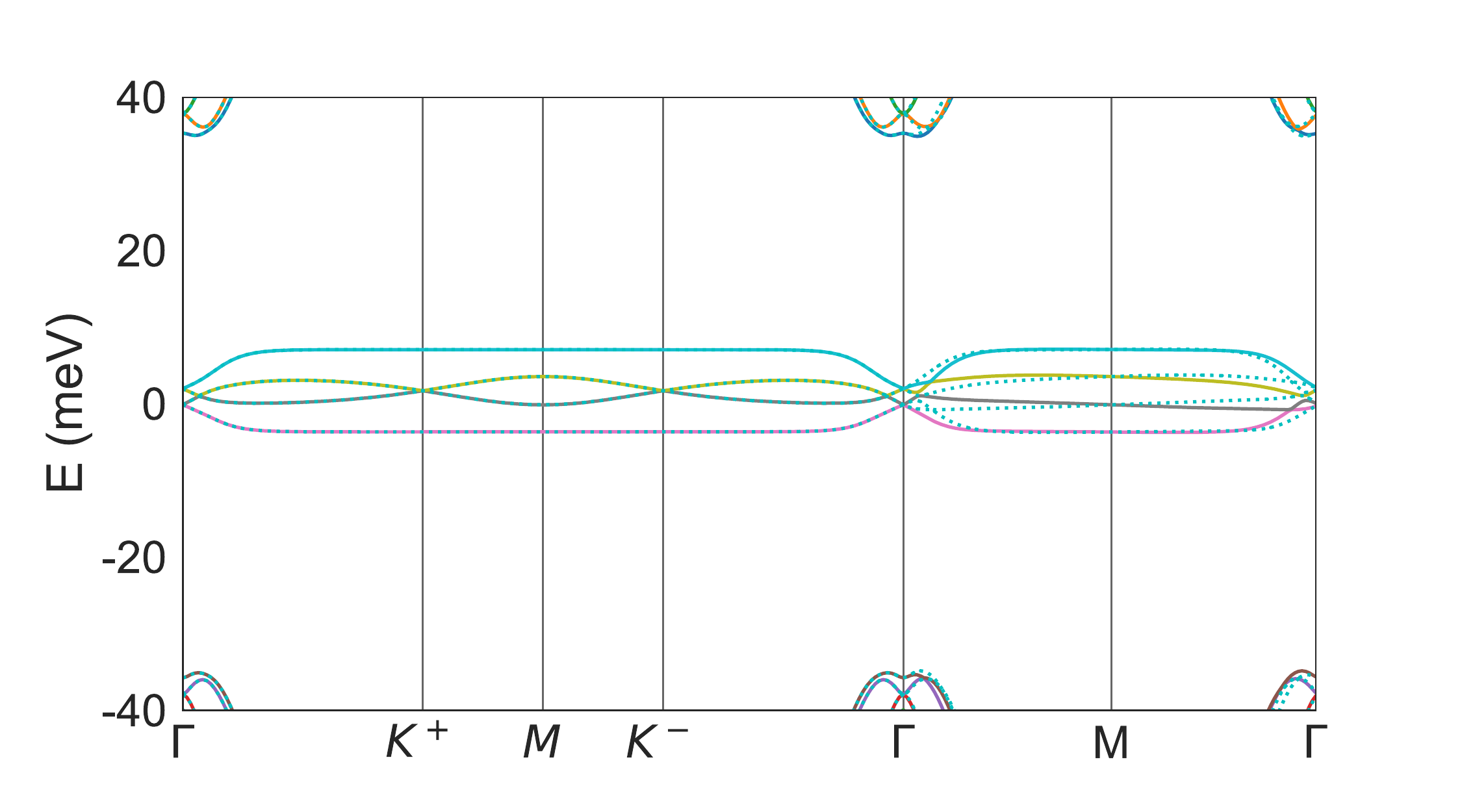}
    \\ (b) \\
    \includegraphics[width=8cm]{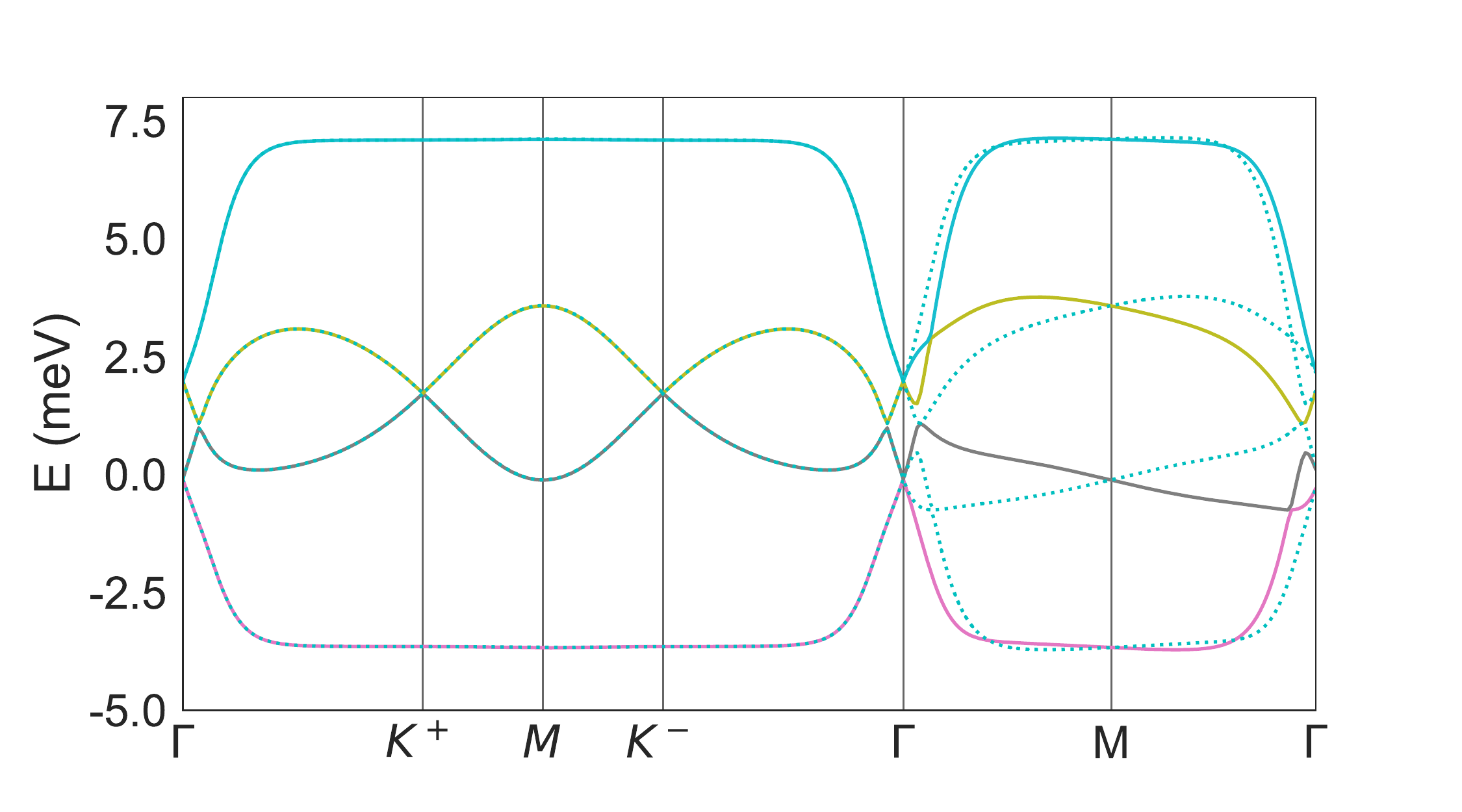}
    \\ (c)
    \caption{(a) Band structure along a high symmetry line of the BM Hamiltonian with a Rashba SOC term given by $\lambda_R^+(\tau^z\sigma^x s^y - \sigma^ys^x)$. The twist angle is $\theta= 1.08$\textdegree, and $\lambda_R^+ = 16$ meV. (b) Same as above, but now with the Rashba SOC as $\lambda_R^-l^z(\tau^z\sigma^x s^y - \sigma^ys^x)$ which has opposite sign on both layers, and $\lambda_R^- = 16$ meV. (c) Same as (b) but zoomed in to highlight the flat bands. The solid and dotted lines represent the flat bands from different valleys.}
    \label{fig:FlatBand}
\end{figure}

There are multiple different ways to combine TBG with TMD substrates, depending on whether a TMD substrate is placed on only one or on both sides of the TBG device, and depending on the in-plane orientation of the TMD relative to the graphene layers. In this work, we consider all these possibilities and find that the different devices have very different band structures. We also find that $\mathcal{C}_{2x}'$, the spinful two-fold rotation symmetry around the in-plane $x$-axis, plays an important role in explaining the differences in band structure. In particular, our results show that TBG-TMD heterostructures which are $\mathcal{C}_{2x}'$ symmetric have Dirac points near the $\Gamma$ point, even though the spinful $\mathcal{C}_{2z}\mathcal{T}$ symmetry is broken, while $C_{2x}'$ breaking heterostructures have a gapped band spectrum. Interestingly, we find that the gapped band structures have a pair of bands which are extremely flat around the $K^\pm$ point. For a representative band structure displaying this pair of flat bands, see Figs.~\ref{fig:FlatBand}(b) and \ref{fig:FlatBand}(c), where the band structure is shown along a path between high-symmetry points in the mini-Brillouin zone. We develop an intuitive understanding of the spin-orbit coupled band structures, and in particular how they are affected by the different proximity-induced terms, by doing a $k\cdot p$ analysis at the different Dirac points.

For the devices with a gapped band spectrum, we calculate the Chern numbers of the different isolated bands. We find a very rich phase diagram depending on the relative strength of the different proximity-induced terms. We identify the parameter regions and integer filling factors where valley Chern insulators and topological insulators protected by time-reversal symmetry are realized. Interestingly, we find that for realistic values of the proximity-induced terms, it is possible to obtain topological insulators both at filling $\nu = -2$ and $\nu = 2$, i.e. at the filling factors obtained by doping either two holes or two electrons per moir\'e unit cell into the system.

The remainder of this paper is organized as follows. In Section \ref{sec:tbg}, we start by reviewing the continuum model of twisted bilayer graphene. In Section \ref{sec:tmd} we consider the different TBG-TMD heterostructures, and discuss the proximity-induced terms that appear in the TBG Hamiltonian, together with their symmetries. We then first study the effect of only the leading Rashba spin-orbit coupling terms in Section \ref{sec:rashba}. We use a $k\cdot p$ analysis to develop an understanding of the effects of Rashba spin-orbit coupling on the TBG flat bands, and we investigate the symmetry protection of the different Dirac cones in the bands with spin-orbit coupling. In Section \ref{sec:ising}, we add the remaining subleading proximity induced terms, and analyze the resulting flat bands. Again, we resort to a $k\cdot p$ analysis to develop an intuitive understanding of the effects of the subleading terms. Next, we focus on the heterostructures which can have gapped flat bands, and we calculate the corresponding Chern numbers in Section \ref{sec:chern}. We also discuss the different topological phases that can be realized in these devices. We end with a discussion of our results in Section \ref{sec:discussion}.

\section{Twisted bilayer graphene continuum model}\label{sec:tbg}

In this section, we first review the continuum model describing twisted bilayer graphene (TBG) \cite{CastroNeto,Bistritzer11,Morell2010}. We start by defining the single-valley, spinless moir\'e Hamiltonian, which can be written as
\begin{equation}\label{singlevalley}
    H_K = \sum_{\k}H_0(\k) + H_T(\k) \, .
\end{equation}
The first term $H_0(\k)$ is the intra-layer term, given by
\begin{equation}
    H_0(\k)=\sum_{l}f_l^\dag(\k)h_{l\theta/2}(\k)f_l(\k)\, .
\end{equation}
Here, $f_l^\dag(\k)$ denotes the electrons from layer $l$, each having two components from the two sublattices. $l=+/-$ is the index of top/bottom layer, and $h_{\theta}(\k)$ is the monolayer graphene Hamiltonian with twist angle $\theta$ and Fermi velocity $v_F$:
\begin{equation}
    h_{\theta}(\k)=\hbar v_{F}\left(k_{x} \sigma^x+k_{y} \sigma^y\right) e^{-i \theta \sigma^z}\, ,
\end{equation}
where $\sigma$ acts on the electron sublattices. The top and bottom layer in $H_0(\k)$ are rotated by $\pm\theta/2$ respectively, so the two layers have a relative twist angle $\theta$.

The second term in Eq.~\eqref{singlevalley},  $H_T(\k)$, is the inter-layer tunneling term given by
\begin{equation}
	H_T(\k)=\sum_{j=1}^{3}f_+^\dag(\k+\mathbf{q_j})T_j f_-(\k)+h.c.
	\, .
\end{equation}
The momentum transfer $\mathbf{q}_1$ is defined as $\mathbf{q}_{1} = \mathbf{K_-}-\mathbf{K_+}$, i.e. it corresponds to the momentum difference between the Dirac points of the bottom and top layer. The other two momentum transfers $\mathbf{q}_2$ and $\mathbf{q}_3$ are related to $\mathbf{q}_1$ by the three-fold rotation symmetry: $\mathbf{q}_{2} = \mathcal{C}_{3z}\mathbf{q}_{1}$, $\mathbf{q}_{3}= \mathcal{C}_{3z}\mathbf{q}_{2}$. The inter-layer hopping matrices $T_j$ are defined as
\begin{equation}
    T_{j}=w_{0} \sigma^{0}+w_{1}\sigma^x e^{\frac{2\pi i}{3}(j-1)\sigma^z}\,. 
\end{equation}
The two parameters $w_0,w_1$ respectively correspond to the sublattice diagonal (AA/BB) and sublattice off-diagonal (AB/BA) hopping strengths respectively. In this work, we use $w_1=110$ meV and $w_0=0.75w_1$, which takes into account corrugation effects \cite{Nam,Koshino,Kaxiras}. For these values of the inter-layer hopping, the single-valley TBG Hamiltonian has two very flat bands around charge neutrality for twist angles close to the first magic angle value $\theta^*\sim 1.08$\textdegree.

Adding the spin and valley degrees of freedom, the complete moir\'e Hamiltonian for TBG takes the form
\begin{equation}
    H_{TBG} = \left(\begin{matrix} H_K(\k) & \\ & H_{\bar{K}}(\k) \end{matrix}\right)_\tau \otimes s^0\, ,
\end{equation}
where $\tau \in \{+,-\}$ labels the two valleys, and $s^\mu$ correspond to the identity matrix and the three Pauli matrices in spin space. The moir\'e Hamiltonians coming from different valleys are related by time reversal: $H_{\bar{K}}(\k) = H_K^*(-\k)$.

The Hamiltonian $H_{TBG}$ preserves the following symmetries: (1) Time-reversal symmetry $\mathcal{T}$; (2) $\mathcal{C}_{3z}$ rotation around the out-of-plane $z$-axis; (3) $\mathcal{C}_{2z}$ rotation around the $z$-axis and (4) $\mathcal{C}_{2x,2y}$ rotation around the in-plane $x$- and $y$-axes. For the purposes of this work, the most relevant symmetries are $\mathcal{T}$, $\mathcal{C}_{2z}$ and $\mathcal{C}_{2x}$, which respectively act as $\mathcal{T}=\tau^x\mathcal{K}$ ($\mathcal{K}$ is complex conjugation), $\mathcal{C}_{2z}=\tau^x\sigma^x$ and $\mathcal{C}_{2x}= l^x\sigma^x$. Note in particular that both $\mathcal{T}$ and $C_{2z}$ interchange valleys, and are therefore absent in the single-valley moir\'e Hamiltonian $H_K(\k)$. As expected form their geometrical definition, all two-fold rotations interchange the sublattices, and only rotations about in-plane axis interchange the two layers.

$H_0(\k)$ contains two Dirac cones which are coupled in the presence of $H_T(\k)$. As a result, $H_{TBG}(\k)$ has two Dirac cones in each valley with renormalized Fermi velocity \cite{Bistritzer11}. Importantly, the Dirac cones are protected by the $\mathcal{C}_{2z}\mathcal{T}$ symmetry, which acts within a single valley.

\section{Combining TBG with a TMD substrate}\label{sec:tmd}

As is known from previous work \cite{Gmitra15}, placing graphene on a transition metal dichalcogenide (TMD) substrate (such as $\mathrm{MoS_2}$) can induce  significant spin-orbit coupling (SOC) for the graphene electrons. Because the TMD lattice is highly incommensurate with graphene, its effect on the low-energy spectrum of graphene can be well approximated by a spatially-independent perturbation $H_{SOC}$ on the graphene layer nearest to the substrate. The most important SOC terms which are induced are the Ising SOC, which takes the form 
\begin{equation}
H_{SOC_I}=\lambda_I\tau^z s^z\, ,
\end{equation}
and the Rashba SOC, which is given by
\begin{equation}
H_{SOC_R}=\lambda_R(\tau^{z}\sigma^{x}s^{y}-\sigma^{y}s^{x})\, ,
\end{equation}
where $\tau, \sigma, s$ act on the  graphene layer proximate to the TMD. Note that magnitudes of the SOC terms in principle depend on the angle of the TMD-graphene alignment. In particular, the TMD is not invariant under a $\mathcal{C}_{2z}'$ rotation, leading to the $\mathcal{C}_{2z}'$-odd term $H_{SOC_I}$: if the alignment is rotated by 180\textdegree, $\lambda_I \to - \lambda_I$.
In recent experiments, the typical strengths of the different proximity-induced terms in each layer were found to be $\lambda_R \approx 16$ meV, $\lambda_I\approx 1$ meV and $u\approx 1$ meV \cite{Avsar14,Wang15,Gmitra16,Yang16,Island19}.

In magic angle graphene the kinetic energy of the electrons is quenched and the SOC strength becomes comparable to the bandwidth of the nearly flat bands. For this reason, the SOC can potentially have a drastic effect on the moir\'e bands.
We note that the effect of SOC on TBG was also studied previously in Ref.\cite{Harpreet} using the same model studied here, but in the regime below the magic angle ($\theta \sim 0.8$\textdegree).

In addition to the SOC terms, the TMD also induces a finite sublattice splitting, given by
\begin{equation}
    H_{SL} = u \sigma^z\, .
\end{equation}
The sublattice splitting is also induced by an aligned hexagonal Boron-Nitride substrate \cite{Hunt13,Amet13,Zibrov,Jung,YankowitzJung,Kim2018}, and has been observed experimentally to have a non-trivial effect on the correlated phase diagram of TBG \cite{Sharpe,YoungAH,Andreananosquid}.

Because of the SOC, the TBG-TMD heterostructures is no longer invariant under the spinless symmetries mentioned in the previous section. We therefore need to consider the spinful generalizations of the three relevant symmetries discussed above, which are given by $\mathcal{T}'=i\tau^x s^y\mathcal{K}$, $\mathcal{C}'_{2z}=i\tau^x\sigma^x s^z$ and $\mathcal{C}'_{2x}=i l^x\sigma^x s^x$. This will help us determine how the signs of $\lambda_I$, $\lambda_R$ and $u$ depend on the relative orientation of the TMD layer to the TBG. Namely, the $\mathcal{C}'_{2z}$ operation, which is equivalent to a 180\textdegree\  rotation of the TMD layer, anti-commutes with $H_{SOC_I}$ and $H_{SL}$ but commutes with $H_{SOC_R}$. This means that the $\lambda_I$ and $u$ terms will change sign upon changing the TMD orientation, but $\lambda_R$ term will not. We will denote the proximity-induced terms with a $\pm$ superscript, which corresponds to their relative sign on both layers (i.e., $\lambda^+_I$ denotes Ising coupling with the same sign on both layers).

In this work, we focus on three different kinds of TBG-TMD heterostructures: (1) a TMD-TBG-TMD heterostructure with the same TMD orientation on both sides (``even''), (2) a TMD-TBG-TMD heterostructure with opposite TMD orientation on both sides (``odd''), and (3) a one-sided TBG-TMD heterostructure. The even/odd heterostructures add even/odd proximity-induced terms to the TBG Hamiltonian, respectively given by
\begin{align}
    H^{PI}_e &= \frac{u^+}{2}\sigma^z+\frac{\lambda^+_I}{2}\tau^z s^z+\frac{\lambda^-_R}{2}l^z(\tau^{z}\sigma^{x}s^{y}-\sigma^{y}s^{x}) \label{eq:He}\, ,\\
	H^{PI}_o &= l^z\left(\frac{u^-}{2}\sigma^z+\frac{\lambda^-_I}{2}\tau^z s^z+\frac{\lambda^-_R}{2}(\tau^{z}\sigma^{x}s^{y}-\sigma^{y}s^{x})\right)\, . \label{eq:Ho}
\end{align}
Note that the Rashba term is required to have opposite sign on the two graphene layers by the layer-exchanging $\mathcal{C}'_{2x}$ symmetry. The one-sided TBG-TMD stacking only has proximity-induced terms on one layer, and can be written as a combination of the even and odd proximity-induced terms:
\begin{equation}
    H^{PI}_{1s} = \frac{1 + l^z}{2}\left(\frac{u}{2}\sigma^z + \frac{\lambda_I}{2}\tau^z s^z + \frac{\lambda_R}{2}(\tau^{z}\sigma^{x}s^{y}-\sigma^{y}s^{x})\right)\, . \label{eq:Hs}
\end{equation}

In Table~\ref{tab:SB}, we list all the proximity-induced terms (Ising SOC, Rashba SOC and sublattice splitting) in both the even and odd form, and their transformation properties under time-reversal and different two-fold rotations. From the table, we see that the different TBG-TMD heterostructures are distinguished by their transformation properties under $C_{2x}'$. In particular, $H_e$ breaks $\mathcal{C}'_{2x}$, while $H_o$ is invariant under it. Importantly, because of the Ising SOC and sublattice splitting, all heterostructures necessarily break not only the spinless $C_{2z}$ symmetry, but also the spinful $C_{2z}'$ symmetry. All proximity-induced terms, however, do respect the spinful time-reversal symmetry $\mathcal{T}'$. As in the original TBG continuum model, the single-valley moir\'e Hamiltonians of the TBG-TMD heterostructures are interchanged by time-reversal symmetry.

\begin{table}[t]
	\begin{ruledtabular}
		\begin{tabular}{c|ccc|cc}
			\phantom{xx}$m$\phantom{xx} & $(\mathcal{T},\mathcal{T}')$ & $(\mathcal{C}_{2z},\mathcal{C}'_{2z})$ & $(\mathcal{C}_{2x},\mathcal{C}'_{2x})$ & $\mathcal{C}'_{2z}\mathcal{T}'$ & $\mathcal{C}'_{2y}\mathcal{T}'$ \\
			\hline
			\hline
			$u^+$ & $(\checkmark,\checkmark)$ & $(\times,\times)$ & $(\times,\times)$ & $\times$ & \checkmark \\
			$\lambda_I^+$ & $(\times,\checkmark)$ & $(\times,\times)$ & $(\checkmark,\times)$ & $\times$ & \checkmark\\
			$\lambda_{R}^+$ & $(\times,\checkmark)$ & $(\times,\checkmark)$ & $(\checkmark,\times)$ & \checkmark & $\times$\\
			\hline
			\hline
			$u^-$ & $(\checkmark,\checkmark)$ & $(\times,\times)$ & $(\checkmark,\checkmark)$ & $\times$ & $\times$\\
			$\lambda_I^-$ & $(\times,\checkmark)$ & $(\times,\times)$ & $(\times,\checkmark)$ & $\times$ & $\times$\\
			$\lambda_{R}^-$ & $(\times,\checkmark)$ & $(\times,\checkmark)$ & $(\times,\checkmark)$ & \checkmark & \checkmark
		\end{tabular}
	\end{ruledtabular}
	\caption{Symmetry transformation properties of the proximity-induced terms in Eqs.~\eqref{eq:He}, \eqref{eq:Ho} and \eqref{eq:Hs}. A ``$\times$'' entry in the table means that the term on that row is odd under the symmetry labeling the column, i.e. the symmetry is broken. A ``$\checkmark$'' entry means that the proximity-induced term is even and therefore symmetry-preserving.}
	\label{tab:SB}
\end{table}

\section{Flat bands from Rashba spin-orbit coupling}\label{sec:rashba}
As mentioned in the last section, Rashba SOC is the dominant term introduced by the proximity effect of TMD. Therefore, we will first isolate its effect on the flat band spectrum of TBG, and later add the Ising SOC and sublattice splitting as small perturbations.

We first consider a TBG-TMD heterostructure with both even and odd Rashba SOC terms:
\begin{equation}
    H_R=H_{TBG}+\frac{1}{2}\left(\lambda_R^+ +\lambda_R^-l^{z}\right)(\tau^{z}\sigma^{x}s^{y}-\sigma^{y}s^{x}) \label{eq:HR}
\end{equation}
$H_R$ features four flat bands in each valley, with opposite valleys related by $\mathcal{T}'$ symmetry. In Fig.~\ref{fig:FlatBand}, we highlight the four flat bands of $H_R$ from the $K$ valley, with both $(\lambda_R^+,\lambda_R^-)=(16,0)$ meV and $(\lambda_R^+,\lambda_R^-)=(0,16)$ meV. The band spectrum with $\lambda_R^- = 0$, as shown in Fig.~\ref{fig:FlatBand}(a), remains almost two-fold degenerate and is not very different from the original BM band spectrum. The heterostructure with $\lambda_R^+=0$, on the other hand, has a very intricate band structure with several remarkable features as shown in Figs.~\ref{fig:FlatBand}(b) and (c). First, near the $K^\pm$ points there is one pair of well-separated bands which are extremely flat, while the other two bands have Dirac cones at $K^\pm$. Second, we find that in total there are sixteen different linear band crossings or Dirac points, all of which are protected by the spinful $\mathcal{C}'_{2z}\mathcal{T}'$ symmetry. Two of the Dirac points are located at the $\Gamma$ point, and they are displaced in energy, i.e. one Dirac point is between the top two bands and one Dirac point is between the bottom two bands. The three different $\Gamma - \mathrm{M}$ lines each contain two linear band crossings, one between the top two bands, and one between the bottom two bands. The six final Dirac points are very close to, but not exactly on, the $\Gamma - K^+$ and $\Gamma - K^-$ lines. For the parameter values we used, we find that the Dirac points along the $\Gamma- \mathrm{M}$ and $\Gamma -K^\pm$ are all located very closely to the $\Gamma$ point. Because of the spin-orbit coupling, the two-band single-valley BM model becomes a four-band model. As a result, the chirality of the Dirac points is no longer well defined, and it is more appropriate to think about the Dirac points as carrying a non-Abelian charge \cite{SWAhn,WuSoluyanov}.

Below, we use a $k\cdot p$ analysis around both the $K^\pm$ and $\Gamma$ points to develop an understanding of the drastically different effects of the even and odd Rashba SOC on the BM bands, and analyze the symmetry properties of the Dirac points along the $\Gamma-\mathrm{M}$ lines.

Before diving into the details, the fundamental difference between  $\lambda^{-}_R$ [Fig.~\ref{fig:FlatBand}(b)] and $\lambda^{+}_R$ [Fig.~\ref{fig:FlatBand}(c)] can be summarized as follows. While far from the magic angle the two mini-Dirac points at $K^{\pm}$ are localized onto one-or the other layer, near the magic angle the interlayer-tunneling causes the flat bands to be delocalized 50-50 between the two layers. And as explained in more detail below, we also find the flat band states to have a relative phase difference of $i$ between the two layers. As a result, when projecting $H_R$ into the flat band, one finds that the effect of $\lambda^{+}_R$ on the two layers cancels, leaving the band structure almost unaffected, while for $\lambda^{-}_R$ they add. The flatness of the band throughout most of the mBZ is simply a result of the momentum-independence of the Rashba coupling.

\subsection{Analysis of the Dirac points at $K^\pm$}

For our $k\cdot p$ analysis near the $K^\pm$ points, we start by considering the approximation of the ''chiral limit'' of tBLG, which corresponds to artificially putting $w_0 = 0$ in the BM Hamiltonian \cite{Guinea,Tarnopolsky19}. In the chiral limit, the Bloch states of the flat bands around both the $K^\pm$ points can be taken to be completely localized on one of the two sublattices (i.e $A$ vs $B$). Therefore, we can write the four-component Bloch states of the spinless, single-valley BM Hamiltonian in this sublattice-polarized basis as $\psi(\mathbf{r})$ and $\chi(\mathbf{r})$, where both $\psi$ and $\chi$ have two components corresponding to the two different layers, but live on different sublattices. The chiral Bloch states for the complete moir\'e Hamiltonian are then given by $\psi_{\k}(\mathbf{r})$, $\chi_{\k}(\mathbf{r})$, $\psi_{\mathbf{\bar{k}}}(\mathbf{r})$ and $\chi_{\mathbf{\bar{k}}}(\mathbf{r})$ , where $\{\k, \bar{\mathbf{k}}\}$ denote the $K/\bar{K}$ valleys, and two spins are represented by the same Bloch states. 

We can use the symmetries of the BM model to relate the different chiral Bloch states. First, $H_{TBG}$ is invariant under $\mathcal{C}_{2z}\mathcal{T}=\sigma^x \mathcal{K}$ symmetry, which exchanges the sublattices. Therefore the Bloch states of $H_{TBG}$ which live on different sublattices transform into each other by $\mathcal{C}_{2z}\mathcal{T}$, which implies that $\chi_{\k}(\mathbf{r})=\psi^*_{\k}(-\mathbf{r})$. Similarly, $H_{TBG}$ is also invariant under $\mathcal{T}=\tau^x \mathcal{K}$ symmetry, giving the following relation between other basis functions: $\{\psi_{\mathbf{\bar{k}}}(\mathbf{r}),\chi_{\mathbf{\bar{k}}}(\mathbf{r})\}=\{\psi^*_{-\k}(\mathbf{r}),\chi^*_{-\k}(\mathbf{r})\}$. These relations allow us to fix all chiral Bloch states given a single Bloch state $\psi_{\k}(\mathbf{r})$.

For the purpose of our $k\cdot p$ analysis, we are only interested in the chiral Bloch state $\psi_{\k}(\mathbf{r})$ near the $K^\pm$ points. We have observed numerically that in this region the chiral Bloch states are to a very good approximation given by the following simple form:
\begin{equation}
    \psi_{K^\pm}(\mathbf{r})
	\sim
	\left(
	\begin{array}{c}
	{1} \\ {-i}
	\end{array}
	\right)f(\r)\, ,\label{simpleform}
\end{equation}
where $f(\r)$ is some envelope function which ensures that the wavefunction amplitude is concentrated on the AA regions. Eq.~\eqref{simpleform} thus implies that the wavefunction components on different layers have the same magnitude and a relative phase difference of $\pi/2$. Importantly, we find that away from the chiral limit, and even for realistic values of $w_0$, one can still find a basis for the flat bands where the expression in Eq.~\eqref{simpleform} remains a very good approximation for the basis states. This basis precisely corresponds to the sublattice polarized basis introduced in Ref. \cite{KIVC}. The observation that we can work in a basis of the form in Eq.~\eqref{simpleform} at realistic values of $w_0$ will help us understand many of the results obtained below.

Now we explain how we construct the $k \cdot p$ Hamiltonian near the $K^\pm$ points. We start with numerically diagonalizing $H_{TBG}$ at the $K^\pm$ points. Because the bands at the $K^\pm$ are degenerate, the corresponding eigenbasis is not uniquely defined. To remedy this, we fix the eigenbasis by imposing that the $\mathcal{C}'_{2z}\mathcal{T}'$ symmetry acts as $i\sigma^x s^x\mathcal{K}$. Next, we project $H_R(\mathbf{k})$ in the neighborhood of the $K^\pm$ points into the symmetry-fixed eigenbasis, and obtain the following matrix:
\begin{align}
    \left(\tilde{H}_{R,K^\pm}(\k)\right)_{ij}\equiv\left\langle i \middle | H_R(\k) \middle | j\right\rangle\, ,
\end{align}
where $\ket{i}\in\{\ket{\psi_{K^\pm}},\ket{\chi_{K^\pm}},\ket{\psi_{\bar{K}^\pm}},\ket{\chi_{\bar{K}^\pm}}\}$.
This procedure gives us the following effective $k\cdot p$ Hamiltonian to first order in $\mathbf{k}$:
\begin{equation}
    \tilde{H}_{R,K^\pm}(\k)=E_{0,K}+\tilde{v}_{K}(k_x\sigma^x+k_y\tau^z\sigma^y)+\frac{\tilde{\lambda}_R^-}{2}(\tau^{z}\sigma^{x}s^{y}-\sigma^{y}s^{x})\,,\label{projectedRashba}
\end{equation}
where $\tilde{v}_{K}$ is the renormalized Dirac velocity of the tBLG. Note that under the flat band projection, the even Rashba SOC term ($\lambda_R^+$) vanishes, and therefore it has no effect on the band spectrum in first order perturbation theory. The odd Rashba terms $\lambda_R^-$, on the other hand, is unaffected by the projection. One can explicitly check these properties of the even and odd Rashba terms under projection into the flat bands by using the approximate expression for the flat band basis states given in Eq.~\eqref{simpleform}. 

In each valley, the projected odd Rashba SOC in Eq.~\eqref{projectedRashba} has eigenvalues $\{-\tilde{\lambda}_R^-,0,0,\tilde{\lambda}_R^-\}$. This separates two of the four zero-energy bands with quantum number $\sigma^z s^z\tau^z=-1$ with a band gap $\Delta=2\tilde{\lambda}_R^-$, producing the flat bands in Figs.~\ref{fig:FlatBand}(b) and (c), while the other two bands remain unchanged and retain their Dirac crossing. These Dirac points are protected by the $\mathcal{C}'_{2z}\mathcal{T}'$ symmetry of $H_R$.

\subsection{Analysis of the Dirac points on the $\Gamma - M$ lines}

\begin{figure}
    \centering
    \includegraphics[scale=0.7]{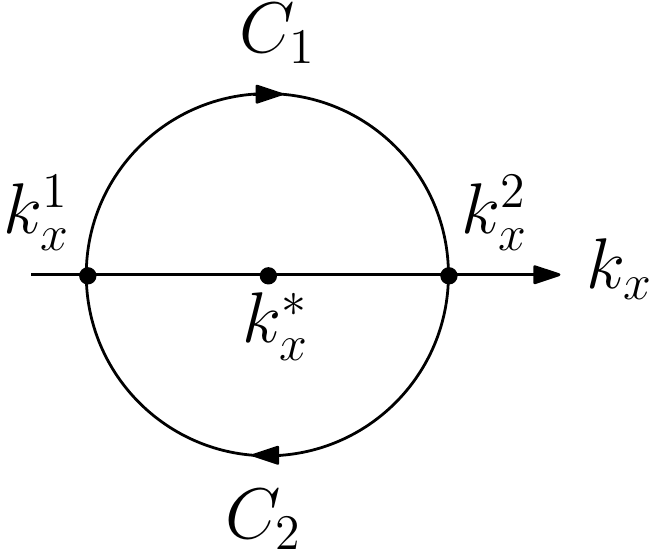}
    \caption{$C_{2x}'$ symmetric contour around a Dirac point at $\mathbf{k}^* = (k_x^*,0)$. The contour consists of two parts $C_1$ and $C_2$, which are related by $C_{2x}'$. $k_x^1$ and $k_x^2$ are the begin (end) and end (begin) points of $C_1$ ($C_2$).}
    \label{fig:contour}
\end{figure}

Having analyzed the Dirac cones at the $K$-points, we now turn our attention to those Dirac cones in the band structure of Fig.~\ref{fig:FlatBand}(c) which lie on the three different $\Gamma - \mathrm{M}$ lines. Our goal is to identify the symmetries which protect these Dirac points. Let us first consider the Dirac cones at the $\Gamma$ point, and focus on the case with $\lambda_R^+=0$. As a first step, we again perform a $k\cdot p$ analysis. We do this by numerically diagonalizing $H_R$ at the $\Gamma$ point, and then using the eigenstates to construct the $k\cdot p$ Hamiltonian. To fix the eigenbasis, we impose different symmetry representations with a new basis $\{\tau, \tilde{\sigma}, \tilde{s}\}$: $\tilde{\mathcal{T}}'=i\tau^y\mathcal{K}$, $\tilde{\mathcal{C}}'_{2z}\tilde{\mathcal{T}}'=\mathcal{K}$, and $\tilde{\mathcal{C}}'_{2x}=i\tau^z\tilde{\sigma}^z \tilde{s}^z$. Here, $\tau$ still refers to the valley degree of freedom, but $\tilde{\sigma}$ and $\tilde{s}$ no longer represent the original sublattices or spins, as they are now hybridized by the Rashba SOC. Instead, $\tilde{\sigma}^z$ distinguishes the pair of high energy ($\tilde{\sigma}^z = +$) and the pair of low energy bands ($\tilde{\sigma}^z = -$) at $\Gamma$, and $\tilde{s}^\mu$ acts within each pair of high or low energy bands. Note that in this basis, the $\tilde{\mathcal{C}}'_{2x}$ operator is diagonal matrix, which will be helpful for our later discussions.

After projecting $H_R(\mathbf{k})$ in the neighborhood of the $\Gamma$ point into this eigenbasis, we obtain the following effective $k\cdot p$ Hamiltonian to first order in $\mathbf{k}$:
\begin{equation}
    \tilde{H}_{R,\Gamma}=\left(E_{0,\Gamma}+\tilde{m} \tilde{\sigma}^z\right) +\tilde{v}_{\Gamma}\tau^{z}(k_x\tilde{s}^{z}+k_y\tilde{\sigma}^{z}\tilde{s}^{x}) \label{eq:kpGamma}
\end{equation}
In each valley, $H_R$ has two separated Dirac cones at the $\Gamma$ point. Similar to the Dirac cones in the BM band spectrum, they are protected by the valley-charge conservation symmetry and $\mathcal{C}'_{2z}\mathcal{T}'$. 

From the $k\cdot p$ analysis we also observe that all valley-diagonal mass terms, i.e. the terms $\tau^{0,z}\tilde{\sigma}^{0,z}\tilde{s}^y$, not only break the $\tilde{\mathcal{C}}'_{2z}\tilde{\mathcal{T}}'$ symmetry, but also the $\tilde{\mathcal{C}}'_{2x}$ symmetry. As a next step, we will show that this is no accident, and that the Dirac cone at $\Gamma$ is protected by the $\mathcal{C}_{2x}'$ symmetry. At the same time, we will also show that the additional Dirac points along the $\mathcal{C}_{2x}'$-invariant $\Gamma-\mathrm{M}$ line are also protected by $\mathcal{C}_{2x}'$, and that the Dirac points along the two $\Gamma - \mathrm{M}$ lines which are interchanged by $\mathcal{C}_{2x}'$ are protected by a combination of the $\mathcal{C}_{2x}'$ and $\mathcal{C}_{3z}'$ symmetries, where the latter is the spinful three-fold in-plane rotation symmetry.

To show that Dirac cones along the $\mathcal{C}_{2x}'$ invariant line are protected by $\mathcal{C}_{2x}'$, let us consider the general situation where there is a Dirac point at momentum $\mathbf{k}^* = (k_x^*,0)$. For generality, we will also allow the $\mathcal{C}'_{2z}\mathcal{T}'$ symmetry to be broken. Denoting the cell-periodic part of the Bloch states as $|u_{\k,n}\rangle$, the $\mathcal{C}_{2x}'$ symmetry implies that

\begin{equation}
    D(\mathcal{C}_{2x}')|u_{\k,n}\rangle = e^{i\alpha_n(\k)}|u_{\mathcal{C}_{2x}\k,n}\rangle\, ,\label{C2xBloch}
\end{equation}
where $\mathcal{C}_{2x}\k = (k_x,-k_y)$, $D(\mathcal{C}_{2x}')$ is the $\mathcal{C}_{2x}'$ representation acting on the periodic part of the Bloch states, and $e^{i\alpha_n(\k)}$ is a gauge-dependent phase factor. Note that we always work in a continuous gauge, such that $\alpha_n(\k)$ is a continuous function of momentum. Since $\mathcal{C}_{2x}'^2 = -1$, the phase factors $e^{i\alpha_n(\k)}$ satisfy $\mathrm{exp}(i[\alpha_n(\k)+\alpha_n(\mathcal{C}_{2x}\k)])=-1$. Along the $\mathcal{C}_{2x}$-invariant momentum line, the phase factors $\mathrm{exp}(i\alpha_n(\k)) = \pm i$ are the $\mathcal{C}_{2x}'$ eigenvalues of the Bloch states. From the $\mathcal{C}_{2x}'$ transformation property in Eq.~\eqref{C2xBloch}, it follows that the Berry connection $\A_n(\k) = i\langle u_{\k,n}|\mathbf{\nabla}|u_{n,\k}\rangle$ satisfies

\begin{figure*}[ht]
\begin{center}
    \includegraphics[width=14cm]{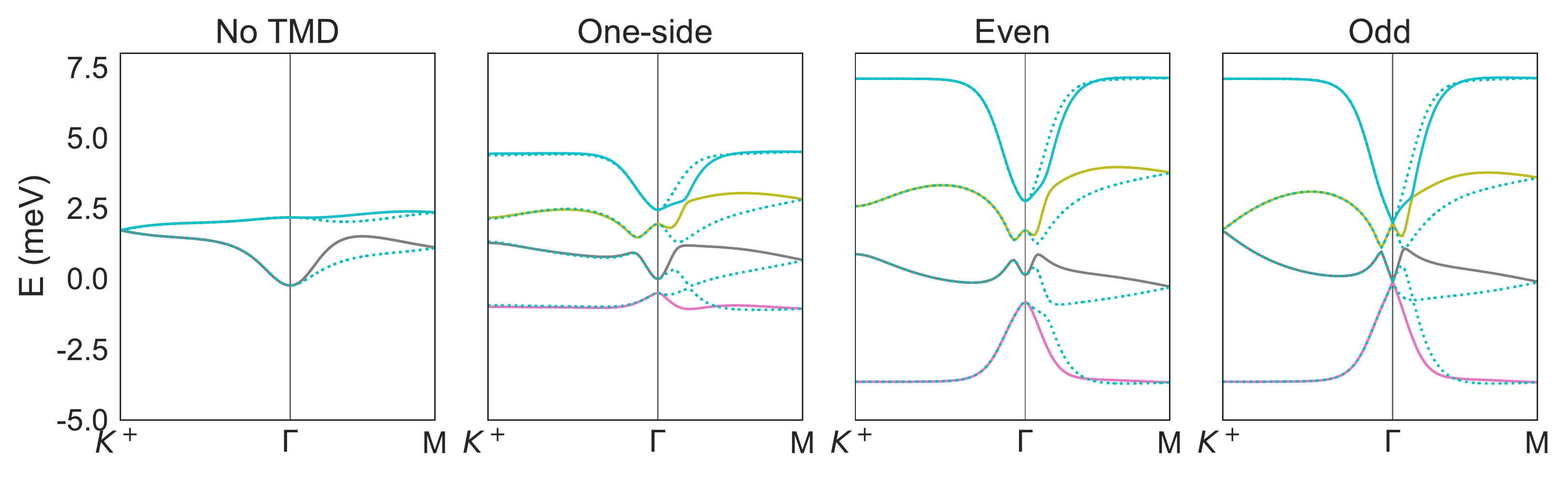}\\
    \hspace{1.05cm}(a)\hspace{1.38cm} \hspace{1.38cm}(b)\hspace{1.38cm} \hspace{1.38cm}(c)\hspace{1.38cm} \hspace{1.38cm}(d)
    \caption{Example of band spectra along $\Gamma-K^+$ line and $\Gamma-M$ line in four different kinds of devices: (a) TBG withouth TMD substrate; (b) one-sided TBG-TMD heterostructures; (c) even TMD-TBG-TMD heterostructures; (d) odd TMD-TBG-TMD heterostructures. The strengths of proximity-induced terms are $(\lambda_R,\lambda_I,u)=(16,1,2)$ for all figures.}
    \label{fig:SandwichBand}
\end{center}
\end{figure*}

\begin{equation}
    \A_n(\k) = \mathcal{C}_{2x}\A_n(\mathcal{C}_{2x}\k) - \mathbf{\nabla}\alpha_n(\k)\, .\label{C2xBerry}
\end{equation}
Note that Eq.~\eqref{C2xBerry} implies that the Berry curvature $F(\k) = \partial_x A_{n,y}(\k) - \partial_y A_{n,x}(\k)$ is odd under $\mathcal{C}_{2x}'$.

Next, we consider a $\mathcal{C}_{2x}$ symmetric contour $C$ encircling the Dirac point at $\k^*$. This contour consists of two parts $C_1$ and $C_2$, which are related by $\mathcal{C}_{2x}$. The two parts $C_1$ and $C_2$ meet at the points $k_x^1$ and $k_x^2$ on the $\mathcal{C}_{2x}$-invariant line $k_y=0$. See Fig.~\ref{fig:contour} for an example of the contour $C$. The Berry phase along the closed contour $C$ is given by

\begin{equation}
\oint_C \A_n(\k)\cdot \mathbf{dk} = \int_{C_1} \A_n(\k)\cdot \mathbf{dk} + \int_{C_2} \A_n(\k)\cdot \mathbf{dk}\label{Berryph}
\end{equation}
Using Eq.~\eqref{C2xBerry}, we can write the contribution to the Berry phase of the $C_2$ section of the contour as

\begin{eqnarray}
\int_{C_2} \A_n(\k)\cdot \mathbf{dk} & = & -\int_{\mathcal{C}_{2x}C_1} \A_n(\k)\cdot \mathbf{dk} \\
 & = & -\int_{C_1} \mathcal{C}_{2x}\A_n(\mathcal{C}_{2x}\k) \cdot \mathbf{dk} \\
 & = & -\int_{C_1} (\A_n(\k)+\mathbf{\nabla}\alpha_n(\k)) \cdot \mathbf{dk}\, ,\label{C1}
\end{eqnarray}
where the minus sign in the first line comes from the fact that $C_2$ and $\mathcal{C}_{2x}C_1$ have different orientations, and in the third line we have used Eq.~\eqref{C2xBerry}.

Combining Eqs.~\eqref{Berryph} and \eqref{C1}, we find that the Berry phase is given by
\begin{equation}
    \oint_C \A_n(\k)\cdot \mathbf{dk} = \alpha_n(k_x^1,0) - \alpha_n(k_x^2,0)\, .\label{C2xEig}
\end{equation}
Because the Berry curvature is odd under $\mathcal{C}_{2x}'$, and because the contour $C$ is $\mathcal{C}_{2x}$ symmetric, we know that the only contribution to the Berry phase can come from the Dirac cone at $\k^*$. This means that the Berry phase is equal to $\pi$. From Eq.~\eqref{C2xEig}, it then follows that the $\mathcal{C}_{2x}'$ eigenvalues of the Bloch states along the $\mathcal{C}_{2x}$-invariant line have to change sign on crossing the Dirac point. This implies that every band has to cross an even number of Dirac points along the $\mathcal{C}_{2x}$-invariant momentum line. In a gapped band spectrum, the $\mathcal{C}_{2x}'$ eigenvalue of the Bloch states has to be constant along the $\mathcal{C}_{2x}$-invariant line $k_y=0$, so Eq.~\eqref{C2xEig} implies that the Dirac cones are protected by the $\mathcal{C}_{2x}'$ symmetry. This can also be understood by noting that since the Berry curvature is odd under $\mathcal{C}_{2x}'$, the Berry phase along any $\mathcal{C}_{2x}$-symmetric contour in a gapped band spectrum has to vanish, while this Berry phase is equal to $\pi$ on encircling the Dirac point. 

The arguments above show that the Dirac points along the $\mathcal{C}_{2x}'$-invariant $\Gamma - \mathrm{M}$ line are protected by the $\mathcal{C}_{2x}'$ symmetry. The $\mathcal{C}_{3z}'$ symmetry then implies that also the Dirac points along the two $\Gamma -\mathrm{M}$ lines interchanged by $\mathcal{C}_{2x}'$ cannot be gapped, which means that these Dirac crossing are protected by the combination of the $\mathcal{C}_{2x}'$ and $\mathcal{C}_{3z}'$ symmetries.

We can explicitly check the above conclusions by revisiting the $k\cdot p$ Hamiltonian in Eq.~\eqref{eq:kpGamma}. The $k\cdot p$ Hamiltonian along the $k_y=0$ line takes the form $\tilde{H}_{R,\Gamma}(k_y=0) = E_{0,\Gamma}+\tilde{m}\tilde{\sigma}^z + \tilde{v}_\Gamma k_x \tau^z\tilde{s}^z$, and the $\mathcal{C}_{2x}'$ symmetry acts as $i\tau^z\tilde{\sigma}^z\tilde{s}^z$. From these expressions, it immediately follows that each band will indeed have opposite $\mathcal{C}_{2x}'$ eigenvalues at different sides of the Dirac points at $\Gamma$. We have also numerically computed the $\mathcal{C}_{2x}'$ eigenvalues of the four flat bands along the entire $\Gamma-M$ line, and we observed that the four bands change their $\mathcal{C}_{2x}'$ eigenvalue not only at $\Gamma$, but also on crossing the other Dirac points on the $\Gamma - \mathrm{M}$ line. This confirms that these Dirac points are indeed protected by the $\mathcal{C}_{2x}'$ symmetry.

\section{Effects of sublattice splitting and Ising SOC}\label{sec:ising}

After our analysis of the effect of Rashba SOC on the BM bands, we will now include the Ising SOC and sublattice splitting terms in our analysis. Specifically, we consider the even and odd TBG-TMD heterostructure, where the Ising SOC ($\lambda_I^\pm$) and sublattice splitting ($u^\pm$) are introduced according to Eqs.~\eqref{eq:He} and \eqref{eq:Ho} respectively:
\begin{equation}
    H_{e,o} = H_{TBG}+H^{PI}_{e,o}\, ,
\end{equation}
and they can be treated as perturbations to $H_R$ in Eq.~\eqref{eq:HR}, with $\lambda_R^+=0$ and $\lambda_R^-\neq0$. A similar treatment can be applied to one-sided TBG-TMD heterostructure
\begin{equation}
    H_{1s} = H_{TBG}+H^{PI}_{1s}\, ,
\end{equation}
but the conclusion would be similar, as we found in the previous section that the $\lambda_R^+$ term in $H_{1s}$ has only a small effect.

Similar to Rashba SOC, the additional proximity-induced terms in $H_e$ and $H_o$ will change the band structure of $H_{TBG}$. In Fig.~\ref{fig:SandwichBand}, we show the band spectra of $H_e$, $H_o$ and $H_{1s}$ along high-symmetry lines. We see that all the band spectra are gapped at the $K^\pm$ points where the gapless Dirac cones are broken by the proximity-induced terms, but the band gap in $H_{e,1s}$ are much larger than that of $H_o$. As for the $\Gamma$ point, $H_{e,1s}$ and $H_o$ also show different characters: $H_o$ still retains the two gapless Dirac cones along the $\Gamma- \mathrm{M}$ line, while $H_{e,1s}$ have all of these Dirac cones gapped out. The three Hamiltonians also gap out the Dirac cones away from high-symmetry lines, but the difference between the resulting spectra is less significant. Overall, it is expected that $H_{1s}$ has a band spectrum similar to $H_e$ since they share the same perturbation terms, and we will now explain the different band features of $H_e$ and $H_o$ in the perspective of both symmetry protection and $k\cdot p$ analysis.

We can understand the different band gaps in $H_e$ and $H_o$ from symmetry considerations. According to Table~\ref{tab:SB}, both the $u^\pm$ and $\lambda_I^\pm$ terms in $H_{e,o}$ break the $\mathcal{C}'_{2z}\mathcal{T}'$ symmetry, removing the protection on the Dirac cones at the $K^\pm$ points. However, the $u^-$ and $\lambda_I^-$ terms in $H_o$ preserve the $\mathcal{C}'_{2x}$ symmetry protecting the Dirac cones along the $\Gamma-M$ line, therefore they will not induce a band gap. The $u^+$ and $\lambda_I^+$ terms, on the other hand, do not preserve $\mathcal{C}'_{2x}$, so we do expect those Dirac cones to be gapped in $H_e$, but not in $H_o$.

\begin{figure*}[t]
    \begin{tabular}{ccc}
         \includegraphics[scale=0.4]{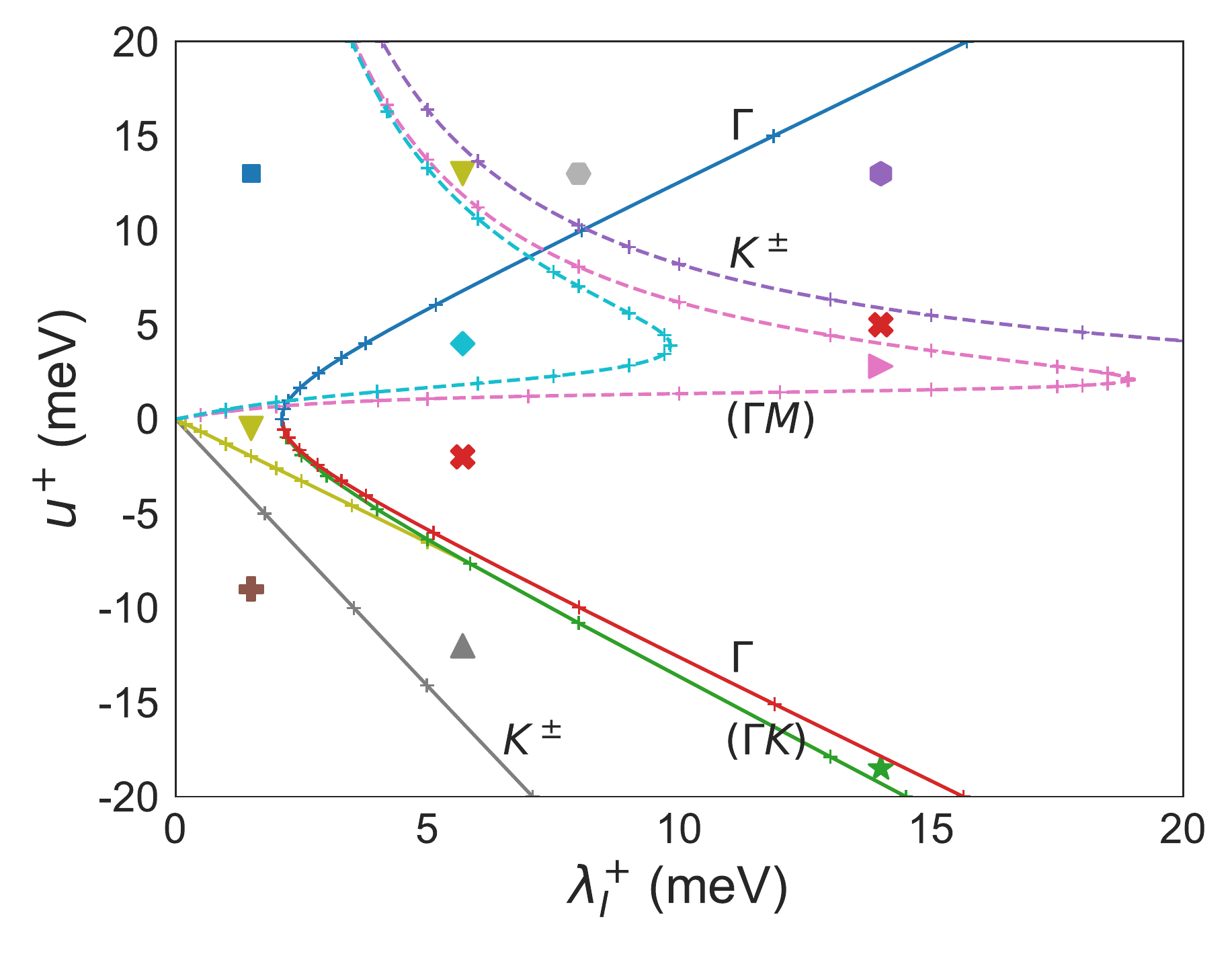} & 
         \includegraphics[scale=0.4]{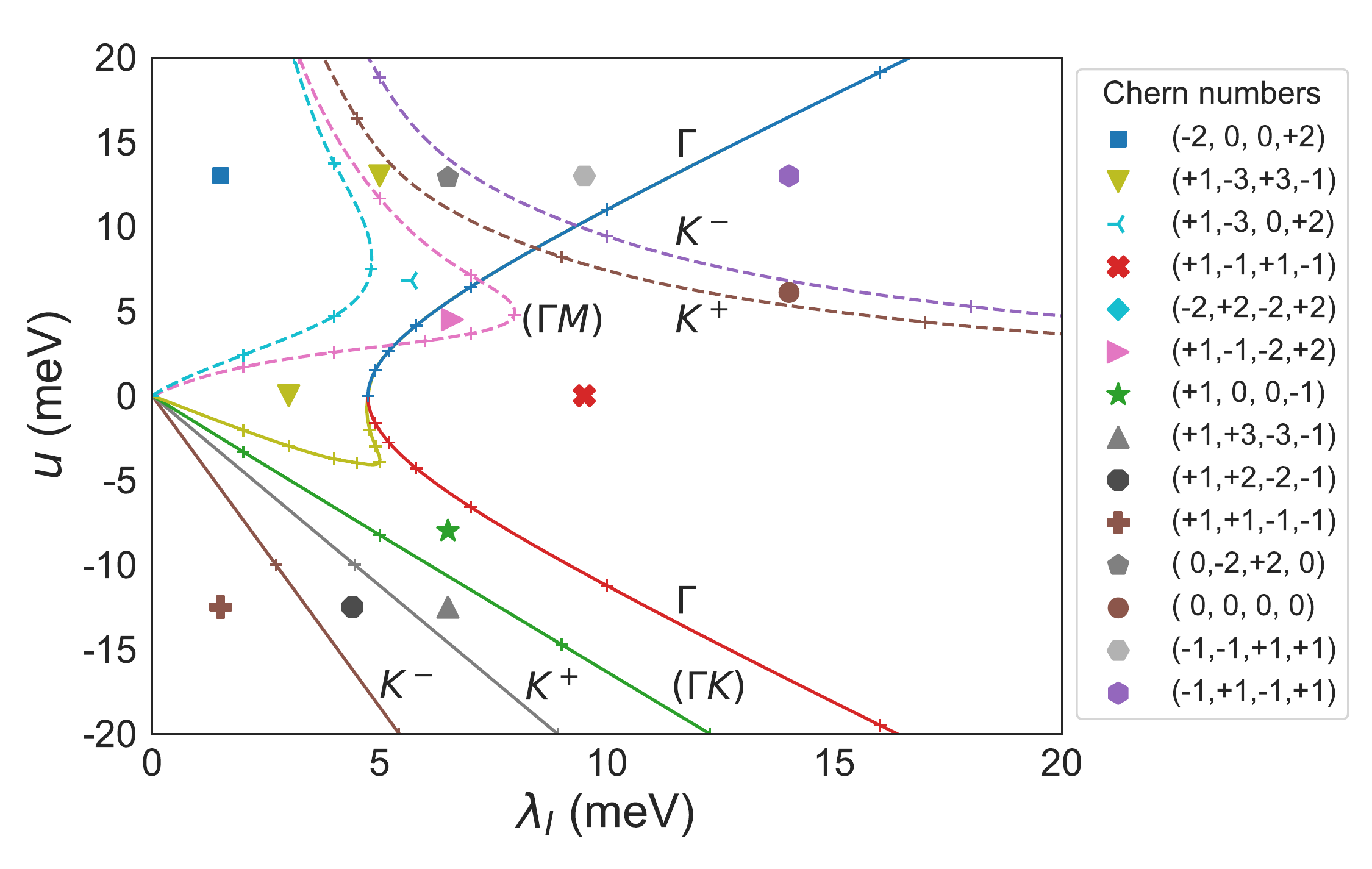} \\
         (a) & (b)
    \end{tabular}
    \caption{Topological phases diagram of (a) even and (b) one-sided TBG-TMD heterostructures regarding $(u,\lambda_I)$ parameters. Each phase is characterized by the Chern numbers of the four flat bands in $K$ valley shown on the right. The sketched phase boundaries are where the band gap vanishes, either between the middle two bands (solid) or the top/bottom two bands (dashed), and the labels represent the position of gap closing in the momentum space.}
    \label{fig:Chern}
\end{figure*}

\begin{figure*}[t]
\begin{center}
	\includegraphics[width=5.5cm]{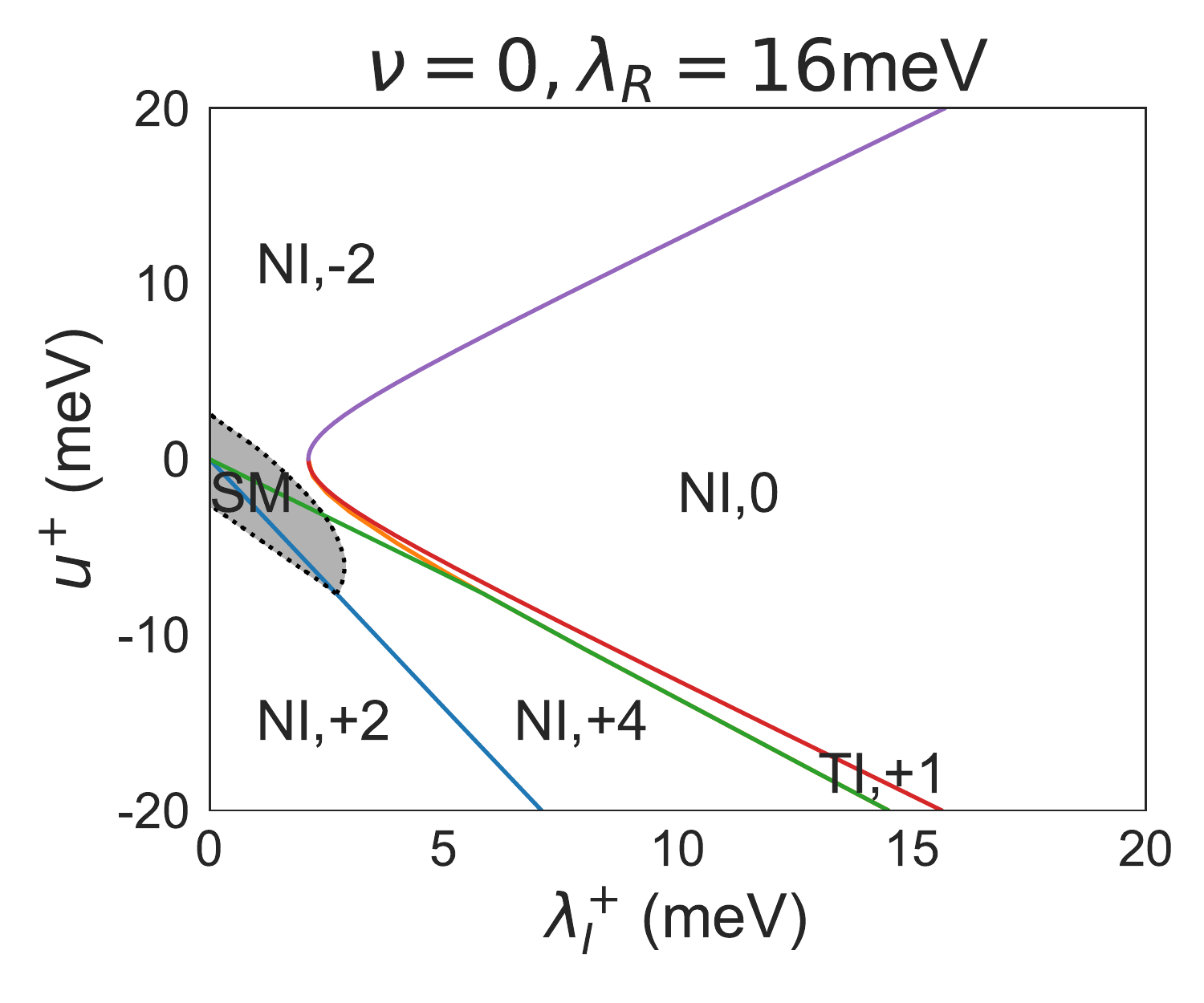}
	\includegraphics[width=5.5cm]{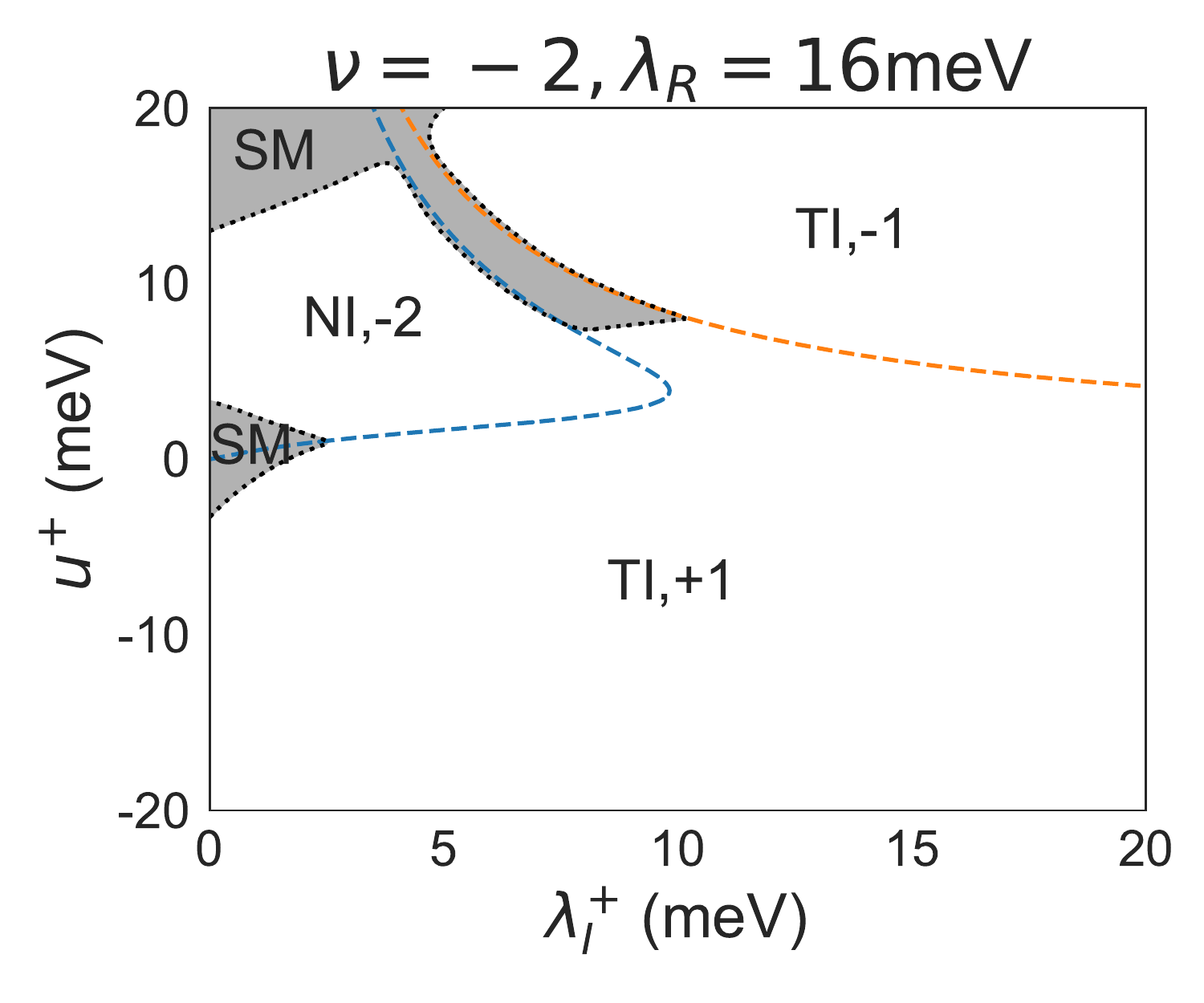} 
	\includegraphics[width=5.5cm]{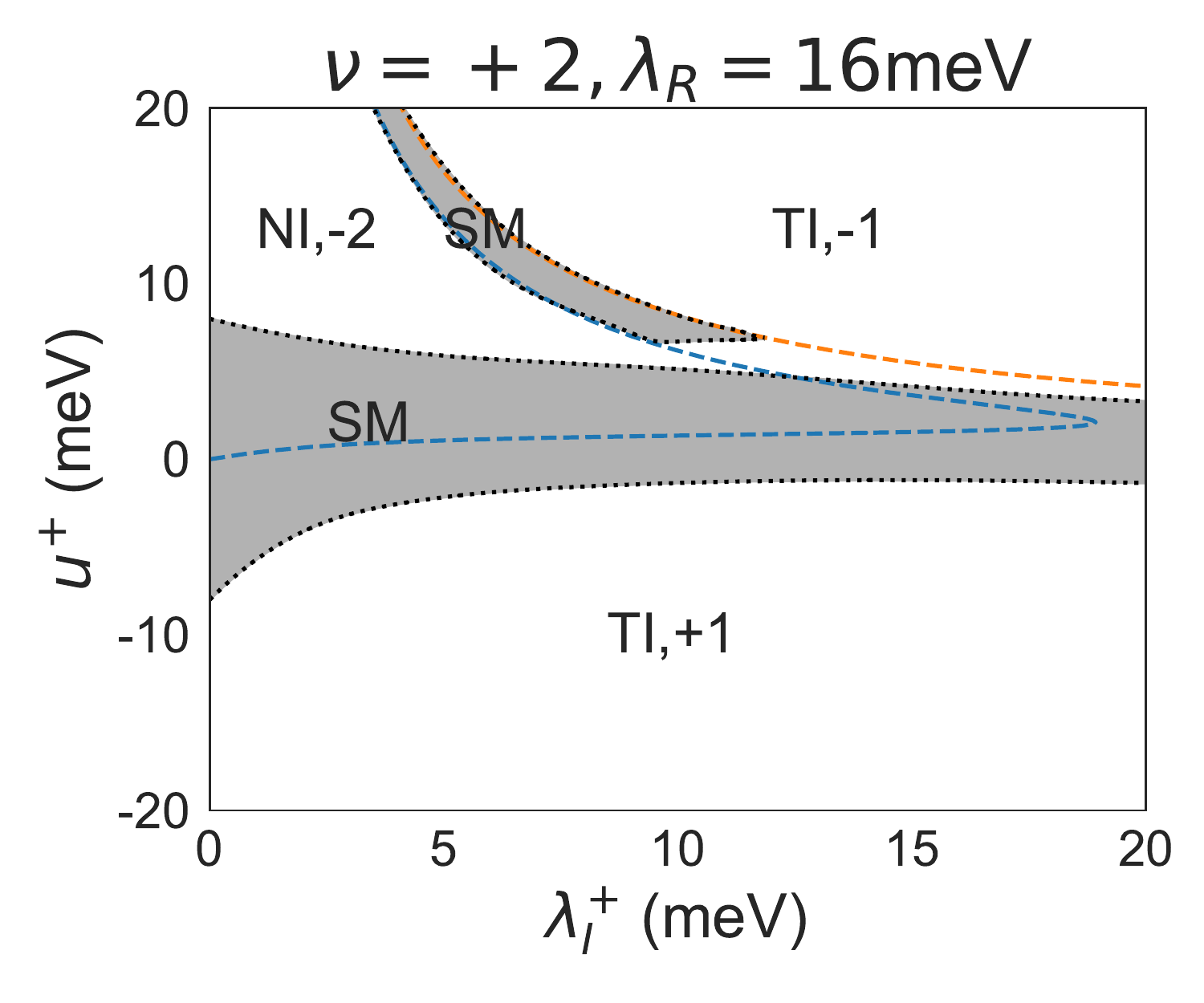} 
	\caption{The topological phase diagram of an even-aligned TBG-TMD heterostructure at $\nu=0$, $\nu=-2$ and $\nu=+2$ filling. Each phase is labeled by the Chern number from the occupied bands in $K$ valley, and categorized into semimetal (SM), normal insulator (NI), and topological insulator (TI) by the $\mathbb{Z}_2$ classification of TI and whether the phase is gapped or not.}
	\label{fig:TIPhase}
\end{center}
\end{figure*}

The arguments above can be made explicit by using our $k\cdot p$ analysis at the $K^\pm$ and $\Gamma$ points. At the $K^\pm$ points, we use the same approximate Bloch states as in the previous section to obtain a $k\cdot p$ Hamiltonian. In this way, we obtain the following effective Hamiltonians for $H_{e,o}$ near the $K^\pm$ points:
\begin{align}
    \tilde{H}_{e,K^\pm} &= \tilde{H}_{R,K^\pm} + \tilde{u}^+\sigma^z+\tilde{\lambda}_I^+\tau^z s^z\, \label{Hekp};\\
    \tilde{H}_{o,K^\pm} &= \tilde{H}_{R,K^\pm}\, .
\end{align}
In $H_e$, the additional terms in Eq.~\eqref{Hekp} gap out the Dirac cone, as they indeed break the $\mathcal{C}'_{2z}\mathcal{T}'=i\sigma^x s^x\mathcal{K}$ symmetry. Note that the $\tilde{u}^+$ and $\tilde{\lambda}_I^{+}$ terms commute, which implies that they are competing mass terms. As for $H_o$, we find that $u^-$ and $\lambda_I^-$ simply vanish under the projection involved in the $k\cdot p$ construction. This can easily be checked explicitly by using the simple approximate form for the BM Bloch states around the $K^{\pm}$, as discussed in detail in the previous section. Even so, $u^-$ and $\lambda_I^-$ terms still gap out the Dirac cones as these terms break the $\mathcal{C}'_{2z}\mathcal{T}'$ symmetry. The Dirac mass generated by these two terms, however, will only appear in second order perturbation theory, and will therefore be much smaller than the Dirac masses in $H_e$. From Fig.~\ref{fig:SandwichBand}, we indeed see that the gaps at the $K^\pm$ points in $H_o$ are significantly smaller than the corresponding gaps in $H_e$, which are barely visible on this scale.

At the $\Gamma$ point, the $k\cdot p$ Hamiltonians corresponding to $H_{e,o}$ are
\begin{align}
    \tilde{H}_{e,\Gamma} &= \tilde{H}_{R,\Gamma} + \tilde{u}^+\tau^z\tilde{\sigma}^y\tilde{s}^z +\tilde{m}^+_I\tau^z\tilde{\sigma}^z\tilde{s}^y\, ,\\
	\tilde{H}_{o,\Gamma} &= \tilde{H}_{R,\Gamma} + (\tilde{u}^-+\tilde{m}^-_I)\tau^z\tilde{\sigma}^y\tilde{s}^x\, ,
\end{align}
With the symmetry representations $\tilde{\mathcal{C}}'_{2z}\tilde{\mathcal{T}}'=\mathcal{K}$ and $\tilde{\mathcal{C}}'_{2x}=i\tau^z\tilde{\sigma}^z\tilde{s}^z$, we see that $\tilde{H}_{e,\Gamma}$ breaks both of these symmetries, while $\tilde{H}_{o,\Gamma}$ breaks $\tilde{\mathcal{C}}'_{2z}\tilde{\mathcal{T}}'$ but not $\tilde{\mathcal{C}}'_{2x}$. This means that only $H_e$ lifts all the symmetry protection on the Dirac cones and obtains a fully gapped bands in first order perturbation. On the other hand, $H_o$ will only renormalize but not gap out these Dirac cones, as they are still under the protection of $\mathcal{C}'_{2x}$.

\section{Valley Chern numbers and topological insulators}\label{sec:chern}

In the previous section we found that both for the even and one-sided TBG-TMD heterostructures the flat bands can be fully gapped in the presence of all proximity-induced terms. This allows us to define Chern numbers for these isolated bands and determine their topological properties. In this section, we examine how these Chern numbers depend on the proximity-induced terms, and what topological phases can be realized in different parameter regimes and for different fillings.

We obtain the Chern numbers for the different isolated bands, as well as for different heterostructures, as follows. First, we note that since all heterostructures preserve time-reversal symmetry $\mathcal{T}'$, we only need to calculate the Chern numbers for the four flat bands in a single valley, as the bands in different valleys related by $\mathcal{T}'$ will have opposite Chern number. The Chern numbers of the isolated bands in one of the valleys are computed using the method of Ref.\cite{Fukui05}. 

Since the band gaps are mainly determined by the strengths of Ising SOC $\lambda_I$ and sublattice splitting $u$, we fix the Rashba SOC strength $\lambda_R =16$ meV, and focus on how the Chern numbers depend on both $u$ and $\lambda_I$. The resulting phase diagrams of the single-valley moir\'e Hamiltonians (in valley $\tau=+$) corresponding to both the even and one-sided heterostructures are shown in Fig.~\ref{fig:Chern}. Note that it is sufficient to consider the parameter regime where $\lambda_I>0$ because the $u^\pm$ and $\lambda_I^\pm$ change sign under $\mathcal{C}'_{2z}\mathcal{T}'$, which implies that the bands with parameters $(-u,-\lambda_I)$ have opposite Chern numbers from those with $(+u,+\lambda_I)$.

From the phase diagrams in Fig.~\ref{fig:Chern}, we can identify the different parameter regimes where, based on a single-particle picture, non-trivial gapped topological phases are realized in the heterostructures. In Fig.~\ref{fig:TIPhase}, we show the possible phases in the even heterostructure at fillings $\nu=-2,0,2$, where bands can fill in $\mathcal{T}'$-related pairs. The different topological phases are characterized by the Chern numbers of the occupied bands, and are distinguished physically in several ways.
The most robust phases are those which have an odd total Chern number in each valley.
These states are topological insulators which are protected by the $\mathcal{T}'$ Kramers symmetry, and are characterized by gapless helical edge modes.
The phases which have an even but non-zero total Chern number in each valley are ``valley-Hall'' states protected only by the valley-charge conservation symmetry. In the bulk, the valley-charge conservation symmetry is preserved to a very good approximation, but it can be significantly broken along the edge of the device. Therefore, the insulators with an even but non-zero valley Chern number are separated from both the trivial and topological insulators by a bulk gap closing, even though their edge modes will acquire a small mass because of the valley-U$(1)$ breaking terms on the edge (while we do not attempt this analysis, these phases are presumably classified by different symmetry-protected distinctions between atomic insulators). It is reasonable to expect the gap of the edge modes is smaller than the bulk gap, such that transport experiments on a device with non-zero even valley Chern number will measure a significantly smaller gap compared to devices with zero valley Chern number.

We can summarize the most relevant experimental implications of our results as follows. First, for the even and one-sided TBG-TMD heterostructure, it is possible to realize a variety of topological phases by tuning the strength of proximity-induced coupling and the filling fraction of the electron flat bands. At filling $\nu=0$, gapped phases can be realized with valley Chern numbers $C_V=\pm2,\pm4$ in two-sided even heterostructures, and also $C_V=\pm1,\pm3$ in one-sided heterostructures. As for filling $\nu=\pm 2$, there is a large and physically realistic parameter regime where phases with $C_V=\pm1$ can be realized, which correspond to non-trivial topological insulators protected by time-reversal symmetry.

\section{Discussion and outlook}\label{sec:discussion}

To summarize, we have analyzed the band structures of different TBG-TMD heterostructures near the magic angle. First, the effect of the dominant Rashba SOC on the TBG flat bands was examined, and the resulting spin-orbit coupled band spectrum was found to exhibit some remarkable features such as a pair of very flat bands near the $K^\pm$ points, and a very high number of Dirac cones. Based on an approximation of the BM Bloch states near the $K^\pm$ point, we provided a simple explanation for how the Rasha SOC gives rise to the very flat bands. We also analyzed the stability of the different Dirac cones, and found that to gap out all Dirac cones one does not only need to break the spinful $\mathcal{C}'_{2z}\mathcal{T}'$ symmetry, but also the $\mathcal{C}_{2x}'$ symmetry, as the latter protects the Dirac cones on the three $\Gamma - \mathrm{M}$ lines (as explained above, some of these Dirac cones additionally require $\mathcal{C}_{3z}'$ to be protected). Having understood the effect of the Rashba SOC, we then included the subleading Ising SOC and sublattice splitting terms. We found that $\mathcal{C}_{2x}'$ breaking heterostructures generically have a fully gapped band spectrum containing eight isolated bands, with the Ising SOC and sublattice splitting terms behaving as competing mass terms driving various phase transitions between different gapped phases. Away from the phase transitions we calculated the Chern numbers of the isolated bands, and we obtained a rich phase diagram with many topologically non-trivial phases. We explicitly identified the parameter regimes where topological insulators and valley-Hall insulators are realized at filling factors $\nu = -2, 0, 2$.

We hope that the analysis presented here can be the starting point for future experimental and theoretical work on TBG-TMD systems near the magic angle. The obvious open question is how the band structures found in this work will be affected by the Coulomb interaction. It would especially be interesting to see how the phase diagram as a function of doping at the magic angle changes in the presence of SOC, as this could potentially provide insight into the nature of the different correlated insulators and the superconducting domes. On the theoretical side, it would be interesting to apply a mean-field analysis and see if the same dominant ordering tendencies are found as in the case without SOC, and whether at the integer fillings there is perhaps a clear candidate symmetry-breaking order which has significantly lower energy. At filling factors $\nu = \pm 2$, it would be especially interesting if no sign of additional symmetry breaking is found in Hartree-Fock, and the topological insulators survive in the presence of interactions, at least on the mean-field level. We leave the study of these questions for future work.

\textit{Acknowledgements --} It is a pleasure  to thank Andrea Young, Ashvin Vishwanath, Eslam Khalaf and Shubhayu Chatterjee for useful discussions. T.W. and M.P.Z. were supported by the Director, Office of Science, Office of Basic Energy Sciences, Materials Sciences and Engineering Division of the U.S. Department of Energy under contract no. DE-AC02-05-CH11231 (van der Waals heterostructures program, KCWF16).

\bibliography{bibliography}

\end{document}